\documentclass[twoside,twocolumn,9pt]{article}
\usepackage{extsizes}
\usepackage[usedoi, usetitle]{rsc} 
\usepackage[version=3]{mhchem}
\usepackage[left=1.5cm, right=1.5cm, top=1.785cm, bottom=2.0cm]{geometry}
\usepackage{balance}
\usepackage{mathptmx}
\usepackage{sectsty}
\usepackage{graphicx} 
\usepackage{lastpage}
\usepackage[format=plain,justification=justified,singlelinecheck=false,font={stretch=1.125,small,sf},labelfont=bf,labelsep=space]{caption}
\usepackage{float}
\usepackage{fancyhdr}
\usepackage{fnpos}
\usepackage[english]{babel}
\addto{\captionsenglish}{%
  
}
\mciteErrorOnUnknownfalse
\usepackage{array}
\usepackage{droidsans}
\usepackage{charter}
\usepackage[T1]{fontenc}
\usepackage[dvipsnames]{xcolor}
\usepackage{setspace}
\usepackage[compact]{titlesec}
\definecolor{darkblue}{rgb}{0,0,0.6}
\usepackage{hyperref}
\hypersetup{colorlinks,linkcolor=darkblue,citecolor=darkblue,urlcolor=darkblue}
\usepackage{amsmath,amssymb}
\usepackage{bm}
\definecolor{cream}{RGB}{222,217,201}

\begin{document}

\pagestyle{fancy}
\thispagestyle{plain}
\fancypagestyle{plain}{
%%%HEADER%%%
\renewcommand{\headrulewidth}{0pt}
}
%%%END OF HEADER%%%

%%%PAGE SETUP - Please do not change any commands within this section%%%
\makeFNbottom
\makeatletter
\renewcommand\LARGE{\@setfontsize\LARGE{15pt}{17}}
\renewcommand\Large{\@setfontsize\Large{12pt}{14}}
\renewcommand\large{\@setfontsize\large{10pt}{12}}
\renewcommand\footnotesize{\@setfontsize\footnotesize{7pt}{10}}
\makeatother

\renewcommand{\thefootnote}{\fnsymbol{footnote}}
\renewcommand\footnoterule{\vspace*{1pt}% 
\color{cream}\hrule width 3.5in height 0.4pt \color{black}\vspace*{5pt}} 
\setcounter{secnumdepth}{5}

\makeatletter 
\renewcommand\@biblabel[1]{#1}            
\renewcommand\@makefntext[1]% 
{\noindent\makebox[0pt][r]{\@thefnmark\,}#1}
\makeatother 
\renewcommand{\figurename}{\small{Fig.}~}
\sectionfont{\sffamily\Large}
\subsectionfont{\normalsize}
\subsubsectionfont{\bf}
\setstretch{1.125} %In particular, please do not alter this line.
\setlength{\skip\footins}{0.8cm}
\setlength{\footnotesep}{0.25cm}
\setlength{\jot}{10pt}
\titlespacing*{\section}{0pt}{4pt}{4pt}
\titlespacing*{\subsection}{0pt}{15pt}{1pt}
%%%END OF PAGE SETUP%%%

\fancypagestyle{SIstyle}{
  \fancyhf{} 
  \renewcommand{\headrulewidth}{0pt}
  \renewcommand{\footrulewidth}{0pt}
  \fancyfoot[C]{\footnotesize\sffamily\thepage}
}

%%%FOOTER%%%
\fancyfoot{}
\fancyfoot[LO,RE]{\vspace{-7.1pt}\includegraphics[height=9pt]{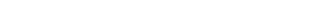}}
\fancyfoot[CO]{\vspace{-7.1pt}\hspace{13.2cm}\includegraphics{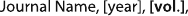}}
\fancyfoot[CE]{\vspace{-7.2pt}\hspace{-14.2cm}\includegraphics{head_foot/RF}}
\fancyfoot[RO]{\footnotesize{\sffamily{1--\pageref{LastPage} ~\textbar  \hspace{2pt}\thepage}}}
\fancyfoot[LE]{\footnotesize{\sffamily{\thepage~\textbar\hspace{3.45cm} 1--\pageref{LastPage}}}}
\fancyhead{}
\renewcommand{\headrulewidth}{0pt} 
\renewcommand{\footrulewidth}{0pt}
\setlength{\arrayrulewidth}{1pt}
\setlength{\columnsep}{6.5mm}
\setlength\bibsep{1pt}
%%%END OF FOOTER%%%

%%%FIGURE SETUP - please do not change any commands within this section%%%
\makeatletter 
\newlength{\figrulesep} 
\setlength{\figrulesep}{0.5\textfloatsep} 

\newcommand{\topfigrule}{\vspace*{-1pt}% 
\noindent{\color{cream}\rule[-\figrulesep]{\columnwidth}{1.5pt}} }

\newcommand{\botfigrule}{\vspace*{-2pt}% 
\noindent{\color{cream}\rule[\figrulesep]{\columnwidth}{1.5pt}} }

\newcommand{\dblfigrule}{\vspace*{-1pt}% 
\noindent{\color{cream}\rule[-\figrulesep]{\textwidth}{1.5pt}} }

\makeatother
%%%END OF FIGURE SETUP%%%

\makeatletter
\newcommand{\onlinecite}[1]{\citenum{#1}}

\makeatother

%%%TITLE, AUTHORS AND ABSTRACT%%%
\twocolumn[
  \begin{@twocolumnfalse}
{\includegraphics[height=30pt]{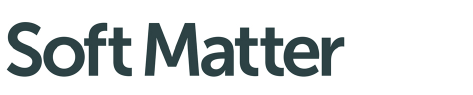}\hfill\raisebox{0pt}[0pt][0pt]{\includegraphics[height=55pt]{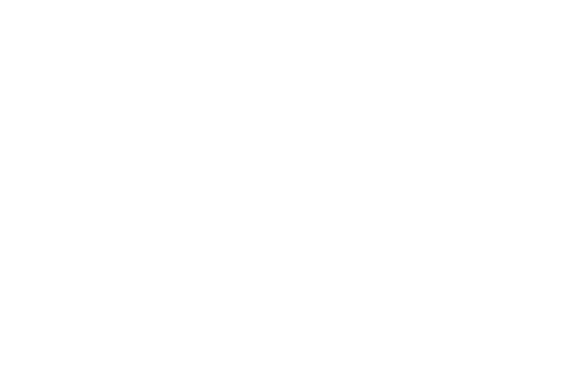}}\\[1ex]
\includegraphics[width=18.5cm]{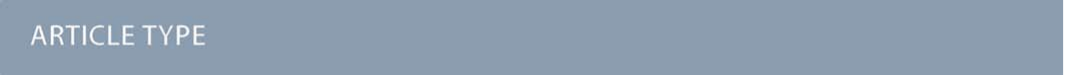}}\par
\vspace{1em}
\sffamily
\begin{tabular}{m{4.5cm} p{13.5cm} }

\includegraphics{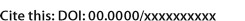} & \noindent\LARGE{\textbf{Self-assembly of quasicrystals under cyclic shear\textsuperscript{\dag}}} \\ 
\vspace{0.3cm} & \vspace{0.3cm} \\

 & \noindent\large{Rapha\"el Maire\textit{$^{*,  \diamond}$}, Andrea Plati\textit{$^{*}$}, Frank Smallenburg\textit{$^{*}$} and Giuseppe Foffi\textit{$^{*,\lozenge}$}} \\

\includegraphics{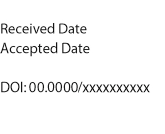} & \noindent\normalsize{We investigate the self-assembly of two-dimensional dodecagonal quasicrystals driven by cyclic shear, effectively replacing thermal fluctuations with plastic rearrangements. Using particles interacting via a smoothed square-shoulder potential, we demonstrate that cyclic shearing drives initially random configurations into ordered quasicrystalline states. The resulting non-equilibrium phase diagram qualitatively mirrors that of thermal equilibrium, exhibiting square, quasicrystalline, and hexagonal phases, as well as phase coexistence. Remarkably, the shear-stabilised quasicrystal appears even where the zero-temperature equilibrium ground state favours square-hexagonal coexistence, suggesting that mechanical driving can stabilise quasicrystalline order in a way analogous to entropic effects in thermal systems. The structural quality of the self-assembled state is maximised near the yielding transition, even though the dynamics are slowest there. Yet, the system still quickly forms monodomain quasicrystals without any complex annealing protocols, unlike at equilibrium, where thermal annealing would be required. Finite-size scaling analysis reveals that global orientational order decays slowly with system size, indicative of quasi-long-range order comparable to equilibrium hexatic phases. Overall, our results establish cyclic shear as an efficient pathway for the self-assembly of complex structures.
}

\end{tabular}

 \end{@twocolumnfalse} \vspace{0.6cm}

  ]

\renewcommand*\rmdefault{bch}\normalfont\upshape
\rmfamily
\section*{}
\vspace{-1cm}

%%%FOOTNOTES%%%

\footnotetext{%
\textit{$^{*}$} Laboratoire de Physique des Solides, Université Paris-Saclay, 510 rue André Rivière, 91400 Orsay, France.\\
\textit{$^\diamond$} \href{https://sites.google.com/view/raphaelmaire}{https://sites.google.com/view/raphaelmaire}\\
\textit{$^\lozenge$} \href{mailto:giuseppe.foffi@universite-paris-saclay.fr}{giuseppe.foffi@universite-paris-saclay.fr}%
}

\footnotetext{\dag~Supplementary Information (SI) available}

%\ddag for both dagguers and reintroduce the part about the SI
%%%END OF FOOTNOTES%%%

\section{Introduction}

From materials science to life itself, the self-assembly of complex structures remains a central challenge in soft matter physics~\cite{whitesides2002self, glotzer2007anisotropy, schrodinger2025life,mcmullen2022self, philp1996self}. While simple periodic crystals are readily formed, the targeted assembly of more intricate structures requires precise control over interparticle interactions and assembly protocols. Among these complex phases, quasicrystals occupy a unique position~\cite{shechtman1984metallic, levine1984quasicrystals}. Characterised by long-range orientational order without translational periodicity, quasicrystals possess forbidden rotational symmetries (such as 5, 8, 10, or 12-fold) and display exotic physical properties, including interesting photonic properties~\cite{Jin1999,zoorob2000complete, florescu2009complete, Vardeny2013} and unusual mechanical properties~\cite{dubois2012properties} such as low friction~\cite{dubois2014friction, silva2016self, yadav2018quasicrystal}.

Since their discovery in metallic alloys, quasicrystalline order has been reported in experiments and simulations in various soft matter systems~\cite{Hayashida2007,barkan2011stability,  Talapin2009, Takano2005, Zeng2004, Zhang2012, Lifshitz2007, Lee2010, Wasio2014, Forster2013, dotera2014mosaic, engel2015computational, pinto2024automating} and, more recently, even in vibrated granular systems~\cite{Plati2024GranularQC}. Reaching these exotic phases through thermal equilibrium is challenging. In soft-matter systems, quasicrystals are typically stable only in narrow regions of the phase diagram, bordered by competing periodic crystals, so their self-assembly often requires targeted approaches such as seeded growth or low-density protocols~\cite{engel2015computational, Noya2021_IQC_directional_bonding, pinto2024automating, kowaguchi2025patchy}. Furthermore, the rugged energy landscape often kinetically traps the system in metastable amorphous states, blocking relaxation to the thermodynamically stable state~\cite{fayen2023self, je2021entropic}. The slow dynamics of colloids (in comparison to atoms or molecules) further exacerbate this issue. Consequently, finding efficient non-thermal alternative pathways to access these complex phases remains an open problem with significant implications for material design and non-equilibrium statistical physics.

A promising approach to circumvent kinetic trapping is the use of mechanical driving, which may provide a means to surpass the constraints imposed by thermal excitations. Recent research has focused intensely on cyclically sheared glasses, where an amorphous configuration is subjected to repeated oscillatory deformation until a steady state is reached. At low strain amplitudes, the system typically settles into an absorbing state, where particles return to the same position after each cycle~\cite{adhikari2018memory}. Conversely, at higher amplitudes, irreversible rearrangements persist, leading to a stroboscopically diffusive steady state~\cite{fiocco2013oscillatory, regev2013onset, Priezjev2013CyclicRelaxation}. Such mechanical driving provides a robust alternative to thermal fluctuations for exploring the potential energy landscape of glasses, allowing the system to either access highly stable configurations or undergo rejuvenation, depending on the strain amplitude~\cite{bhaumik2021role,priezjev2018molecular, schinasi2020annealing, das2022annealing}.

While glasses lack long-range structural order, quasicrystals occupy an intermediate regime between periodic crystals and structurally disordered liquids: they are aperiodic yet exhibit long-range orientational order.  Despite this distinction, glasses and quasicrystals display striking parallels. For instance, both exhibit anomalous vibrational properties~\cite{remenyi2015incommensurate, shintani2008universal, suda2025yielding, bedolla2025striking, mihalkovivc2001atomic, suck1987low}, excess specific heat~\cite{cano2004explanation, cano2004low, zeller1971thermal}, and high configurational entropy relative to crystals~\cite{fayen2024quasicrystal, henley1988random, berthier2019configurational}. Furthermore, they display non-Arrhenius dynamics~\cite{engel2010dynamics, berthier2011theoretical}, dynamical heterogeneities~\cite{bedolla2025relationship,berthier2011dynamical}, and diffusive modes~\cite{landry2020effective, allen1999diffusons, janssen2018aperiodic, jiang2023glassy}, all of which hint at common underlying mechanisms governing their physics~\cite{angell2000amorphous,baggioli2021new, baggioli2019hydrodynamics, baggioli2019universal, baggioli2020unified}.

This motivates asking whether cyclic shear, commonly used to probe and ‘‘train'' glasses, can also promote quasicrystalline order. Simpler systems are known to self-assemble under shear. In particular, crystallisation of simple crystals has been observed under both cyclic shear~\cite{amos2000shear, angelescu2004macroscopic,kerrache2011crystallization, amirifar2019bimodal, haw1998direct, besseling2012oscillatory, koumakis2016amorphous, panaitescu2012nucleation} and monotonic~\cite{blaak2004crystal, daniels2005hysteresis, kerrache2011crystallization, ackerson1990shear, delhommelle2004non, buttinoni2017colloidal} shear. Annealing of grain boundaries~\cite{jana2017irreversibility} or simpler defects~\cite{wang2015cyclic} has also been reported. However, most of these studies involved systems that were not only affected by cyclic shearing but also by thermal fluctuations, dissipative interactions, or hydrodynamic forces. Therefore, we propose to investigate the following question: for a system whose interactions support a stable quasicrystalline phase at equilibrium, can the application of cyclic shear alone drive the system into this phase, especially when simple thermal excitations fail to do so because of kinetic traps?

To address this question, we numerically investigate a two-dimensional particle model known to generate quasicrystalline structures for specific ranges of density and temperature~\cite{fomin2008quasibinary}. By applying both quasi-static and finite shear-rate cyclic deformations, we explore whether such driving can overcome the kinetic limitations inherent to purely thermal evolution. In particular, we examine how the quasicrystalline structure evolves as a function of the deformation amplitude and the number of applied shear cycles.

This work is divided into three main parts. In Sec.~\ref{sec:eq}, we describe the model, which is known to form quasicrystals at equilibrium~\cite{fomin2008quasibinary}. In Sec.~\ref{sec:monotonic}, we apply monotonic shear to the quasicrystal previously formed at equilibrium to familiarise ourselves with the behaviour of quasicrystals under shear. Finally, in Sec.~\ref{sec:cyclic}, we show the self-assembly of quasicrystals under cyclic-shear and investigate its properties.

\section{Numerical model and equilibrium behaviour}\label{sec:eq}

\begin{figure}
    \centering
    \includegraphics[width=0.99\linewidth]{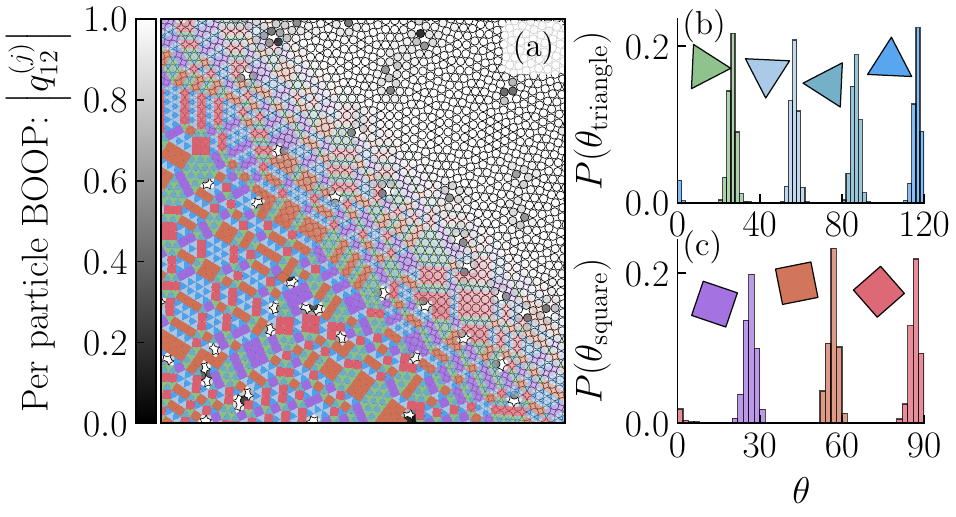}
    \caption{Thermally self-assembled quasicrystal. (a) Part of a typical snapshot with particles coloured by $|q_{12}^{(j)}|$ and the reconstructed tiling. (b) and (c) Orientation distribution of the tiles for the corresponding snapshot. $N = 5000$, $\rho\sigma^2=0.933$ and $k_BT/\varepsilon=0.13$.}
    \label{fig:1}
\end{figure}

\begin{figure}[t]
    \centering
    \includegraphics[width=0.99\linewidth]{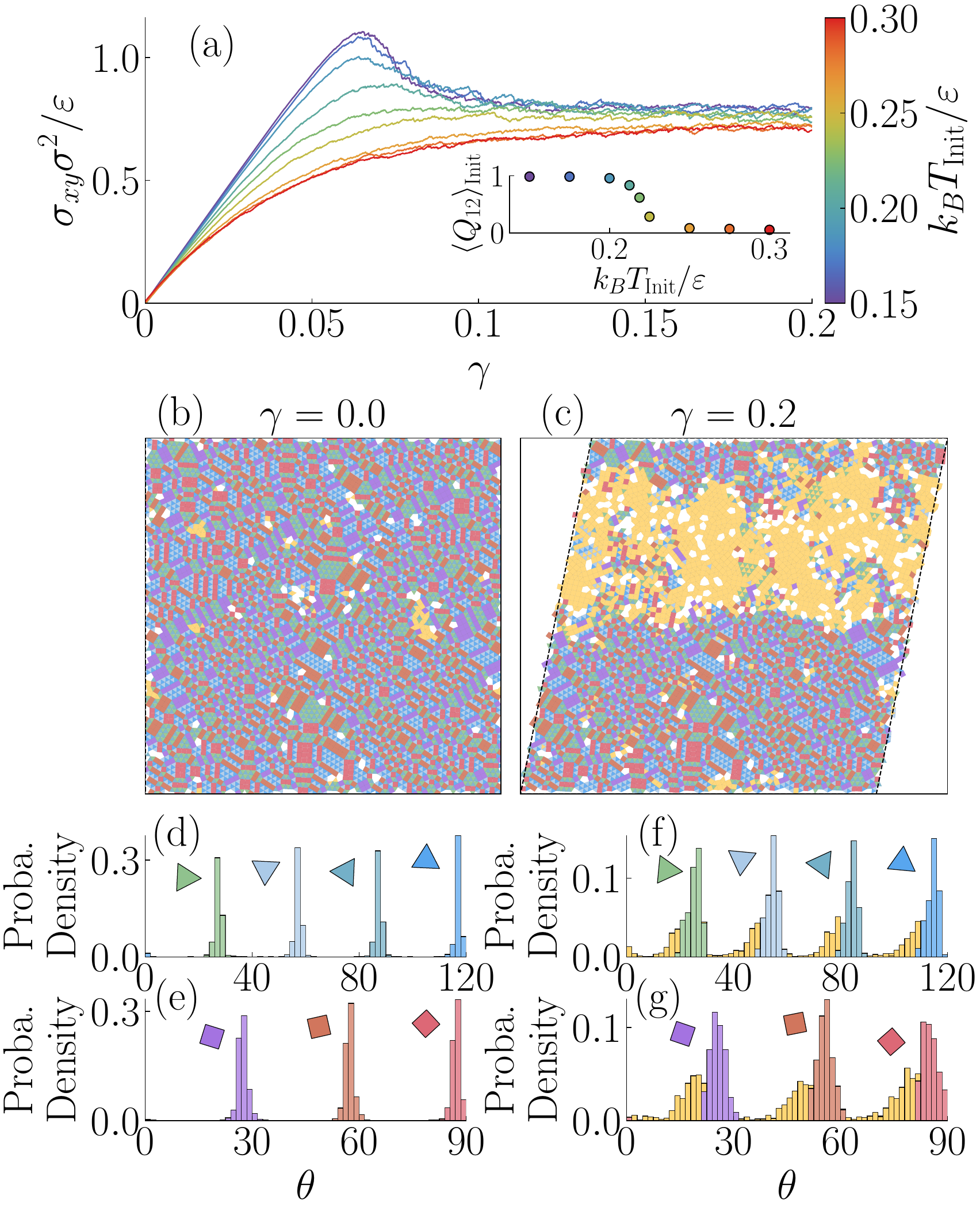}
    \caption{Typical behaviour of the system under monotonic AQS. (a) Stress as a function of strain for various configurations equilibrated at different temperatures. The inset represents the average $Q_{12}$ of the \textit{unstrained} state for different equilibration temperatures. Each curve corresponds to an average over 20 realisations of systems with $N = 10^4$ particles and $\rho\sigma^2 = 0.937$. (b) and (c) show the reconstructed tiling for a system with $N=5\times 10^3$ at $k_BT/\varepsilon=0.2$  for $\gamma = 0$ and $\gamma =0.2$,  respectively. (d, e) and (f, g) show the observed probability distributions of tiles in (b) and (c), respectively. Yellow tiles are tiles considered misaligned with the global orientation of the system, based on an arbitrary angle deviation threshold.} 
    \label{fig:2}
\end{figure}

We consider a two-dimensional system of $N$ particles confined in a periodic box of size $L$. Each particle interacts with its neighbours through a smoothed square-shoulder potential defined as~\cite{fomin2008quasibinary}:
\begin{equation}
    \dfrac{U(r)}{\varepsilon}=\left(\dfrac{\sigma}{r}\right)^{14} + \dfrac{1}{2} \left(1 - \tanh\left[\dfrac{r-\delta}{w}\right]\right),
\end{equation}
where $\varepsilon$ and $\sigma$ set the energy and length scales, respectively. The $\tanh$ term introduces a smoothed shoulder of height $\varepsilon$ centred at $r=\delta$.  $w$ is proportional to the width of the transition region and sets the smoothness of the shoulder. Following previous investigations~\cite{fomin2008quasibinary},  we chose $w =0.1\sigma$ and $\delta=1.35\sigma$. All simulations were performed in \texttt{LAMMPS}~\cite{thompson2022lammps}, with full details provided in the Supplementary Information\textsuperscript{\dag}.

As in hard-disk systems with square shoulders~\cite{dotera2014mosaic,pattabhiraman2015stability,pattabhiraman2017formation,pattabhiraman2017phase,de2022superstripes} and in other models featuring smoothed shoulders~\cite{cardoso2021structural, pattabhiraman2017effect, schoberth2016molecular, somerville2020pattern, gemeinhardt2018growth, malescio2022self, mahynski2017assembly, zhao2024quasicrystals}, this potential promotes quasicrystal self-assembly through the competition between hexagonal packing at high densities $\rho = N/L^2$ and the square-based tilings favoured by the shoulder at lower densities~\cite{kryuchkov2018complex, padilla2020phase, jiang2025mechanical, coli2022inverse, bedolla2025striking}. Such a competition arises in this case by the presence of the two characteristic length-scales $\sigma$ and $\delta$, an effect that can also be obtained with a binary mixture of disks of different size~\cite{fayen2023self}. As an example, in Fig.~\ref{fig:1} we show an equilibrated configuration at $\rho\sigma^2=0.933$ and $k_BT/\varepsilon=0.13$ which displays quasicrystalline order. In Fig.~\ref{fig:1}(a), particles are coloured according to the absolute value of the 12-fold Bond Orientational Order Parameter (BOOP). For particle $j$, the $\ell$-fold BOOP is defined as
\begin{equation}
    q_{\ell}^{(j)}=\dfrac{1}{\mathcal N^{(j)}}\sum_{k\in\mathrm{nn}(j)}e^{i\ell \theta_{jk}},
\end{equation}
where the sum runs over the $\mathcal N^{(j)}$ nearest neighbours $k$ of particle $j$, defined by $|\mathbf r_k - \mathbf r_j|\leq1.35\sigma$ and $\theta_{jk}$ denotes the angle between $\mathbf r_k-\mathbf r_j$ and the $x$ axis. $\left|q_\ell^{(j)}\right|$ is large if the local arrangement of neighbours around the particle $j$ exhibits an $\ell$-fold angular symmetry. We can now define two quantities over the whole system,
\begin{equation}
     q_\ell  =\dfrac{1}{N}\sum_j\left|q_\ell^{(j)}\right|\quad\text{and}\quad Q_\ell = \dfrac{1}{N}\left|\sum_jq_\ell^{(j)}\right|.
     \label{eq:boop}
\end{equation}
The quantity $q_\ell$ reflects the local structure of the system. In contrast, $Q_\ell$ captures global information: in a polycrystalline system, contributions from differently oriented domains partially cancel, driving $Q_\ell\to 0$ even when $q_\ell$ remains large.
For later use, we also define the complex global order parameter
\begin{equation}
\tilde Q_\ell \equiv \frac{1}{N}\sum_j q_\ell^{(j)},
\label{eq:Qtilde}
\end{equation}
so that $Q_\ell = |\tilde Q_\ell|$. The global orientation angle of an $\ell$-fold phase is given by $\arg(\tilde Q_\ell)$.

In addition to BOOPs, we reconstruct the quasicrystalline tiling. The dodecagonal phase decomposes into a perfect planar tiling of equal-edge squares and equilateral triangles, with particles located at the tile vertices (Fig.~\ref{fig:1}(a)). Owing to the quasicrystal’s long-range orientational order, these tiles adopt well-defined favoured orientations, as illustrated in Fig.~\ref{fig:1}(b) and Fig.~\ref{fig:1}(c).

The tiling and the BOOPs contain similar structural information and can both be used to distinguish a randomly oriented square–triangle tiling from a true quasicrystal. For example, an equilibrated amorphous configuration at high density and intermediate temperature may locally resemble a reasonable square–triangle tiling and therefore display a relatively large $\langle q_{12}\rangle$. However, its tile orientations would remain random leading to $\langle Q_{12}\rangle\to0$ since $\text{arg}(q_{12}^{(j)})$ would be broadly distributed. In contrast, a genuine quasicrystal is characterised by a large global order parameter $\langle Q_{12}\rangle$ and therefore, a consistently oriented tiling.

\section{Quasicrystals under monotonic shear}\label{sec:monotonic}

Before looking at the behaviour of the system under cyclic shear, it is useful to look at its behaviour under monotonic shear. We do so using athermal quasi-static shearing~\cite{maloney2006amorphous} (AQS): the system is subjected to a small global strain increment, $\Delta\gamma = 10^{-4}$, applied to the boundaries and particle coordinates, followed by an energy minimisation. The procedure is repeated until the total strain $\gamma$ reaches $\gamma=\gamma_{\max}$.

Similar to many investigations on glasses~\cite{bhaumik2021role,shang2020elastic,jin2023general,ruscher2021avalanches}, before shearing the system, we equilibrate it at a given temperature $T_{\rm init}$ using standard Langevin dynamics, and then use the equilibrated configuration as the initial input for the AQS protocol. The results of our simulations are presented in Fig.~\ref{fig:2}. In panel (a), we show the evolution of the shear stress $\sigma_{xy}$ as $\gamma$ increases, for systems with various initial equilibration temperatures $T_{\rm init}$. When the system is prepared in a high-temperature state, the stress increases smoothly until reaching a plateau, while at a lower temperature, the stress increases to a larger value before decreasing sharply. 

This resembles the ductile-to-brittle crossover in glasses, depending on the degree of annealing~\cite{ozawa2018random, bhaumik2021role, divoux2024ductile}. However, the change in behaviour in our system is most likely also connected to a clear structural change of the system: the formation of a quasicrystal at $T_{\rm init}\lesssim 0.2$. Here we will use the term \textit{brittle} to refer to cases where there is a clear overshoot of the stress around yielding, even if we do not observe, at the present level of annealing, any discontinuity.

In the inset of Fig.~\ref{fig:2}(a), we see that the brittle-like response coincides with the regime where the initial configuration is quasicrystalline, with $\langle Q_{12}\rangle\approx 1$. However, as soon as the system loses its crystallinity or reaches a liquid-quasicrystal coexistence in high-temperature regions, indicated by a drop in $\langle Q_{12}\rangle$, its failure becomes ductile-like. 

Figure~\ref{fig:2}(b,c) shows a quasicrystal configuration and its associated tiling before and after shear; the corresponding tile-orientation distributions are given in Fig.~\ref{fig:2}(d–g). Shearing produces a shear band containing many misaligned tiles (yellow), whereas regions away from the band deform essentially affinely with no plastic rearrangements. Interestingly, tiles within the band are often not fully random: they form a secondary peak at smaller angles in probability distribution, consistent with the quasicrystal blocks sliding above and below the band and rotating it clockwise without fully destroying the pre-existing tiles. This feature is not universal, and other configurations, or larger $\gamma_{\rm max}$, may yield a nearly flat distribution of misaligned tiles, consistent with complete tile breakup during the development of the shear band.

\begin{figure*}[t]
\centering
  \includegraphics[width=0.995\textwidth]{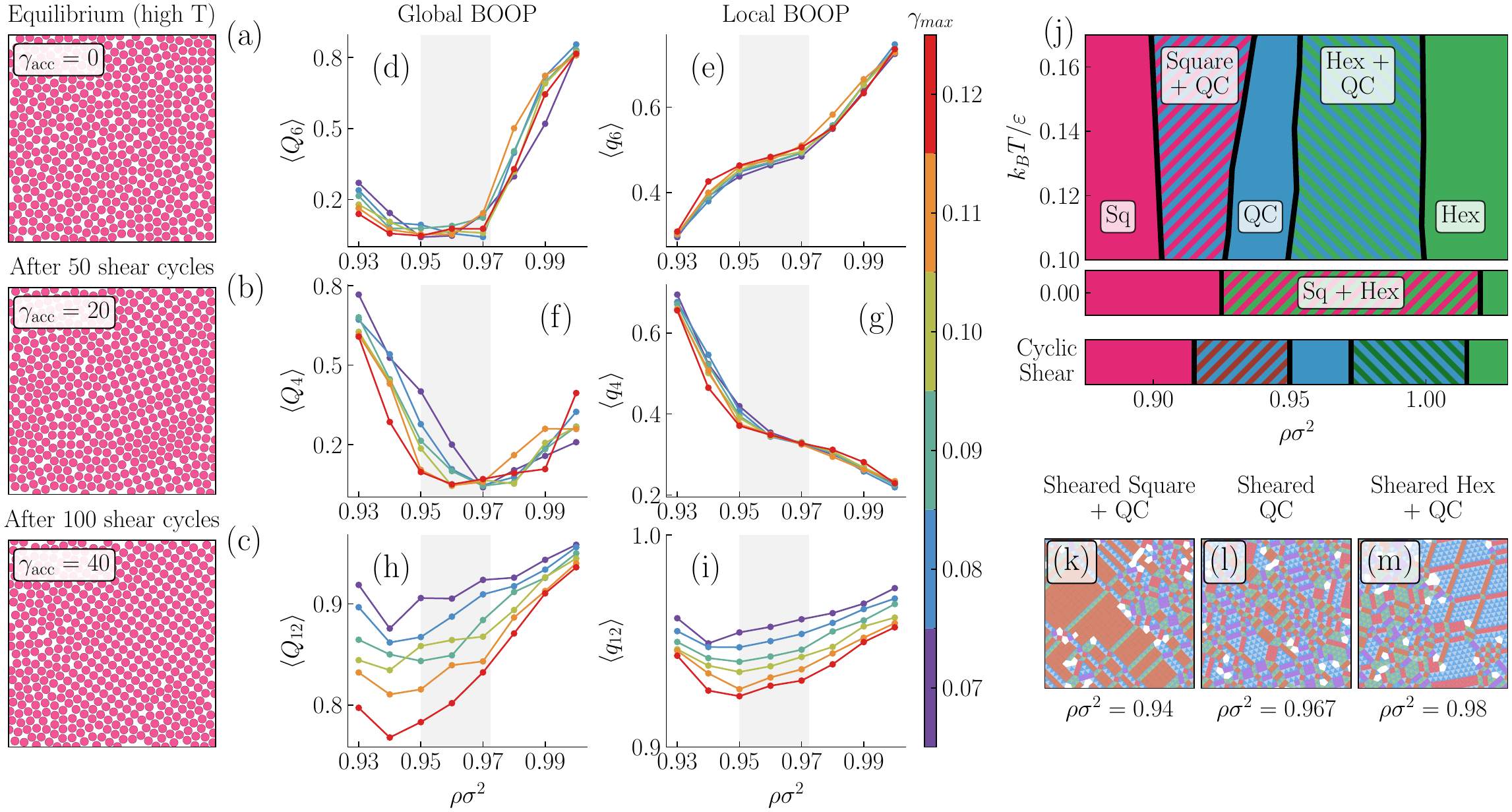}
  \caption{Self-assembly of a crystalline and quasicrystalline structure via cyclic shear. (a-c) Evolution of the system after 0, 50, and 100 cyclic shear cycles at $\gamma_{\max} = 0.1$ and $\rho\sigma^2 = 0.94$ with $\gamma_{\rm acc}=4\gamma_{\max}n_c$, the accumulated strain, where $n_c$ is the number of cycles. The first configuration is an energy-minimized configuration equilibrated at $k_BT/\varepsilon=1$. (d-i) Global and local BOOP (Eq.~\eqref{eq:boop}) as a function of $\rho$ for various $\gamma_{\max}$ in the steady state. The shaded region has quasicrystalline order. The initial configuration is completely random; each point contains an average over 3 different initial configurations and 25 uncorrelated snapshots each. (j) Observed phase diagram. The top part corresponds to the thermal one, reproduced from Refs.~\onlinecite{kryuchkov2018complex} and \onlinecite{padilla2020phase}. The middle part corresponds to the theoretical phase diagram at $T=0$, obtained in the Supplementary Information\textsuperscript{\dag}. The bottom part corresponds to the observed phase in our cyclically sheared system at $\gamma_{\max}=0.1$. The coexistence regions are generally not simply a mixture of the two states at their boundaries, as explained in the main text. (k-m) Typical snapshot at various densities for $\gamma_{\max}=0.1$. All simulations are performed at $N=4000$.}  
  \label{fig:3}
\end{figure*}

\section{Quasicrystal self-assembly via cyclic shear}\label{sec:cyclic}
\subsection{Self-assembly phase diagram}
Much like glasses, which can be trained and structurally reorganised through cyclic shearing above yielding until they reach a stationary state~\cite{fiocco2013oscillatory, regev2013onset, Priezjev2013CyclicRelaxation}, we can similarly study how our system evolves under cyclic shearing, with the possible emergence of a quasicrystalline phase as well as the coexistence between two different phases.

Instead of applying a monotonic shear up to $\gamma = \gamma_{\max}$, the shear direction is now reversed once this value is reached. The system is then sheared to $\gamma = -\gamma_{\max}$ and finally returned to $\gamma = 0$. We define one cycle as
\begin{equation}
    \label{eq:cyc}
    \text{1 Cycle: } \boxed{0}\to\gamma_{\max}\to 0 \to-\gamma_{\max}\to \boxed{0}.
\end{equation}
The system is driven through many such cycles, and its evolution is tracked stroboscopically, that is, configurations and observables are recorded only at the end of each cycle when the strain returns to $\gamma = 0$, i.e., the boxed values in Eq.~\eqref{eq:cyc}.

Starting from a configuration equilibrated at $k_BT/\varepsilon=1$ and at $\rho\sigma^2=0.94$, we apply AQS with $\gamma_{\max} = 0.1$ until a stationary regime is obtained.  Results are shown in Fig.~\ref{fig:3}(a–c). Initially, the system is fully amorphous with no quasicrystalline order. After 50 cycles, however, patches of squares appear, and after 100 cycles, they coalesce into a single square domain coexisting with a quasicrystalline region. We have therefore self-assembled a complex structure via cyclic shearing. It is noteworthy that, at this density and at equilibrium, the system equilibrates to a pure quasicrystalline state for $0.1\lesssim k_BT/\varepsilon\lesssim0.2$, while at lower temperatures it remains essentially kinetically trapped in its initial configuration. 

To elucidate the conditions under which specific structures emerge, we turn to Fig.~\ref{fig:3}(d–i), which displays the behaviour of the BOOPs as a function of density for different values of $\gamma_{\max}$, to reveal the possible structures present in the system in the stationary regime. The left column shows the average global BOOP $\langle Q_\ell \rangle$ ($\ell = 4, 6, 12$) and the right column their local counterparts $ \langle q_\ell\rangle$. At high density, both $\langle q_{6}\rangle$ and $\langle Q_{6}\rangle$ are large, indicating that more compact,  hexagonal patches are dominant. At low density, both $\langle q_{4}\rangle$ and $\langle Q_{4}\rangle$ are large, indicating the presence of less dense energetically favoured square patches. More interestingly, in the intermediate range $0.95\lesssim\rho\sigma^{2}\lesssim0.97$ (shaded in the Figure), both $\langle Q_{4} \rangle$ and $\langle Q_{6}\rangle$ are small while $\langle Q_{12} \rangle$ is large, suggesting the presence of 12-fold quasicrystalline order.

In the same figures, we see that both $\langle q_{12}\rangle$ and $\langle Q_{12}\rangle$ increase as $\gamma_{\max}$ approaches the yielding strain $\gamma_{\rm yield}\simeq 0.065$, which we define as the smallest $\gamma_{\max}$ for which the long-time dynamics is stroboscopically diffusive. This suggests that the shear strength significantly influences the quality of the quasicrystal. Interestingly, the shear amplitude only weakly affects the other structural motifs, with the exception that $\langle Q_{4}\rangle$ increases as $\gamma_{\max}\to\gamma_{\rm yield}$, at low density and in the region where a quasicrystal is formed. This suggests that low-amplitude shear promotes the coherent alignment of square domains with the underlying quasicrystalline order. We will return later to an analysis of the role of $\gamma_{\max}$. For now, we emphasise that near yielding, cyclic shear can generate square, hexagonal, and quasicrystalline structures in the stationary state. Above yielding, the final stationary state is stroboscopically diffusive and therefore independent of the initial configuration, while below yielding, the system evolves into an absorbing state which is highly dependent on the initial configuration~\cite{fiocco2013oscillatory}. For this reason, we do not consider this latter regime.

To place these observations in an equilibrium context, we reproduce in the top panel of Fig.~\ref{fig:3}(j) the equilibrium $\rho$--$T$ phase diagram from Refs.~\onlinecite{kryuchkov2018complex} and \onlinecite{padilla2020phase}. As expected, low densities favour a square lattice, high densities a hexagonal lattice, and an intermediate quasicrystalline phase emerges between them, with coexistence regions separating the phases.

Although sampling the equilibrium states at $T=0$ is impractical, we can obtain the $T=0$ phase diagram (middle of Fig.~\ref{fig:3}(j)) by minimising the energy density of ideal square, triangular, and random tilings\textsuperscript{\dag} (see SI). This analysis shows that the quasicrystal is always energetically disfavoured relative to a coexisting square–hexagonal phase\footnote[3]{This differs from Ref.~\onlinecite{kryuchkov2018complex}, which did not account for phase coexistence when constructing their ground-state phase diagram.}. Thus, at equilibrium, the quasicrystal is stabilised by entropy---likely its configurational component~\cite{fayen2024quasicrystal}---and can only be stable at $T>0$.

Such equilibrium-based arguments cannot \textit{a priori} apply to a system which explores its energy landscape via cyclic shearing rather than thermal motion. Nevertheless, when we map the shear-driven (observed) phases as a function of density for a fixed $\gamma_{\max}=0.1$ (Fig.~\ref{fig:3}j, below the $T=0$ diagram), we find the same qualitative sequence as the density is increased: squares, quasicrystals, hexagons, each separated by coexistence regions. The phase boundaries of the crystals lie roughly between the $T=0$ and $k_{B}T/\varepsilon=0.1$ equilibrium boundaries, suggesting that shear-driven exploration resembles a small but non-zero effective temperature. More interestingly, the quasicrystal is found to self-assemble under cyclic shear, even though it is unstable at equilibrium at $T=0$. Naturally, this interpretation assumes that the energy-minimisation procedure in the AQS protocol explores the phase space only locally, as a non-local minimiser could instead find the true global minimum, which, as shown above, corresponds to square–hexagonal coexistence.  

A naive expectation is that $\gamma_{\max}-\gamma_{\rm yield}$ plays a role analogous to temperature, with larger values promoting particle rearrangements. Our results do not support this picture, at least close to yielding, since even at $\gamma_{\max}=\gamma_{\rm yield}$, the system remains active, with system-spanning avalanches involving a finite fraction of particles. Thus, the limit $\gamma_{\max}\to\gamma_{\rm yield}$ is not analogous to $T\to0$, where activity would vanish. Consistent with this, we find that quasicrystalline order does not decrease as $\gamma_{\max}\to\gamma_{\rm yield}$, unlike the suppression expected as $T\to0$ in equilibrium.

Several other observations lead us to conclude that a thermal analogy does not hold. First, when the system is sheared beyond yielding, the stress–strain curve exhibits hysteresis~\cite{bhaumik2021role}. As a consequence, the structure obtained under cyclic shear at the end of a cycle retains a residual stress $\sigma_{xy}(\gamma = 0)\neq 0$, unlike equilibrium systems. This residual stress indicates that cyclic shear does not sample the same ensemble of potential-energy minima as thermal fluctuations, even though it can still produce equilibrium-like structures.
Second, the region in which the quasicrystal is stabilised by shear is shifted to higher densities compared with the equilibrium system at any temperature.  Finally, the shear-induced coexistences are not simple mixtures of the pure phases found at their boundaries. Fig.~\ref{fig:3}(k) shows that the "square" portion of a square-quasicrystal coexistence contains multiple two-particle-wide hexagonal bands. These bands form consistently at $\pm45^\circ$ to the shear direction and preserve the square-phase orientation. Similarly, the hexagonal region in a hexagonal-quasicrystal coexistence contains two-particle wide square bands with a well-defined orientation, as can be seen in Fig.~\ref{fig:3}(m). Such mixed triangle-square motifs are energetically easy to form, though not as favourable as a mixture of pure phases. The stripe-like features~\cite{imperor2021square} likely form to relieve shear stress by enabling domain sliding. Sliding can arrest at intermediate lattice registries that are local energy minima, yielding square motifs within hexagonal crystals or hexagonal motifs within square crystals. The stripe density within this coexistence regime appears to depend on the overall density and decreases as one approaches either pure crystal. Although a systematic finite-size study would be required to quantify this trend, the resulting coexistence does not resemble a simple lever-rule mixture of the two bounding phases~\cite{olmsted1997coexistence,coclite2014pattern,hu2023direct,xu2003phase}.

\subsection{Varying driving and optimising the quasicrystal}

\begin{figure*}
    \includegraphics[width=0.99\textwidth]{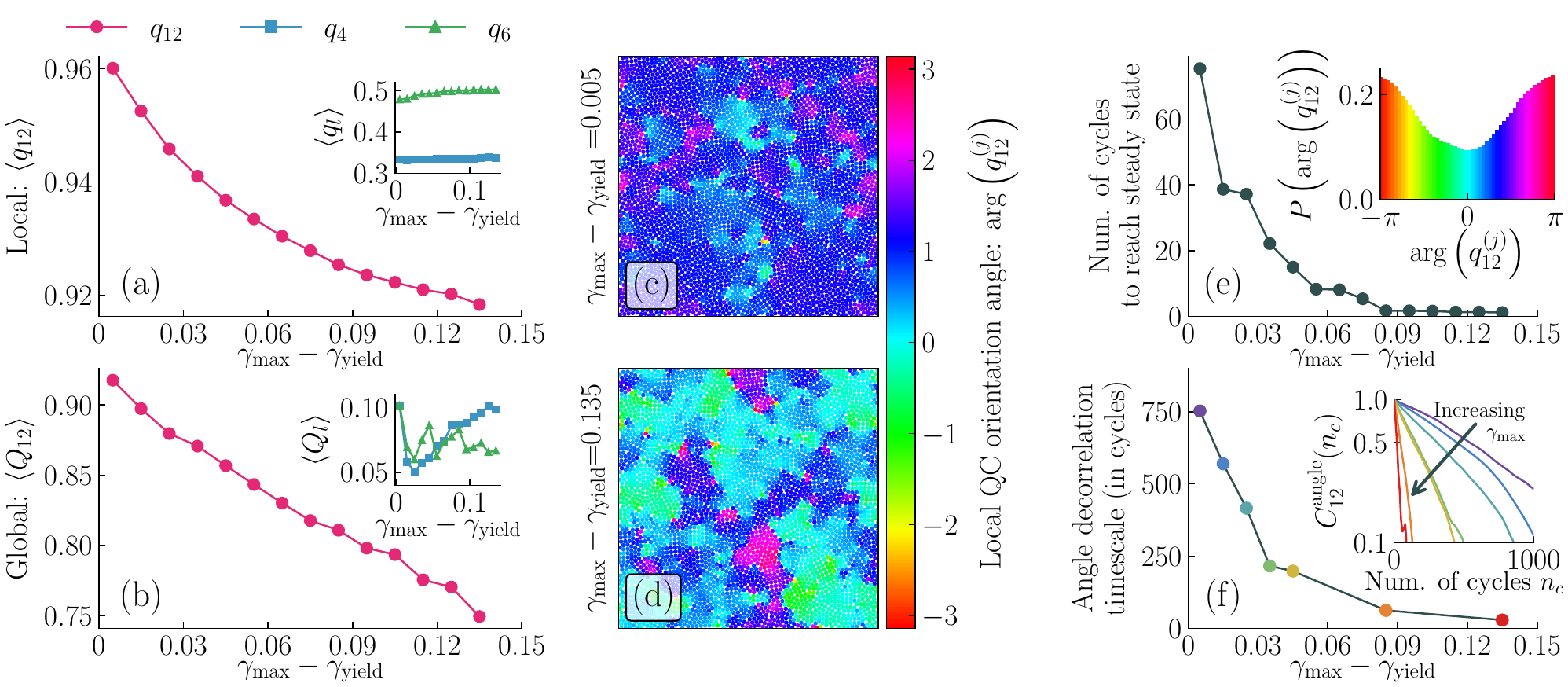}
  \caption{Dependence of the quality of the self-assembled quasicrystal on $\gamma_{\max}$. All simulations were performed at $N=4000$ and $\rho\sigma^2=0.966$. (a–b) Local and global bond-orientational order parameters (Eq.~\eqref{eq:boop}) as a function of $\gamma_{\max}$. Each data point is an average over 10 independent initial conditions and 50 uncorrelated snapshots. (c–d) Typical snapshots at $\gamma_{\max}-\gamma_{\rm yield}=0.005$ (first point in (a) and (b)) and $\gamma_{\max}-\gamma_{\rm yield}=0.135$ (last point in (a) and (b)), respectively, with particles coloured by their local 12-fold orientation. (e) Typical number of cycles required to reach a steady state from a fully random configuration as a function of $\gamma_{\max}$. This self-assembly time is determined by averaging $\langle Q_{12}\rangle(\text{cycles})$ over 10 independent initial conditions and finding the first cycle at which $\langle Q_{12}\rangle(n_c)$ enters and remains within 10\% of its long-time average. The inset shows the probability distribution of the 12-fold orientation angle. (f) Characteristic angle decorrelation timescale (in cycles), obtained from an exponential fit of the 12-fold orientation autocorrelation function $C_{12}^{\text{angle}}(n_c)\sim e^{-n_c/\tau}$ (Eq.~\eqref{eq:C(t)}). Each autocorrelation curve is averaged over at least 5000 configurations sampled every 10 cycles.}
  \label{fig:4}
\end{figure*}

In this section, we examine in detail the role of the driving amplitude in optimising the quality of the quasicrystalline structures. To this end, we fix the density at $\rho\sigma^{2}=0.966$, near the centre of the stability region of the cyclically sheared pure quasicrystal, and vary $\gamma_{\max}$. We focus on the regime $\gamma_{\max}>\gamma_{\rm yield}$, since otherwise the system evolves into an absorbing state. The main findings are summarised in Fig.~\ref{fig:4}.

In Fig.~\ref{fig:4}(a), we show that  the local 12-fold order $\langle q_{12}\rangle$ increases monotonically as $\gamma_{\max}$ is decreased. The inset reveals concurrent decreases of the local 4 and 6-fold parameters, $\langle q_4\rangle$ and $\langle q_6\rangle$, indicating fewer local square and hexagonal patches and a shift toward a more regular tiling with reduced defect density. Fig.~\ref{fig:4}(b) reports the global BOOPs. $\langle Q_{12}\rangle$ likewise increases as $\gamma_{\max}$ approaches yielding, indicating a better-defined global orientation. Interestingly, both $\langle Q_{6}\rangle$ and $\langle Q_{4}\rangle$ show a modest rise near yielding, implying that residual square and hexagonal patches tend to align coherently across the quasicrystalline sample in that regime.

Snapshots in Fig.~\ref{fig:4}(c) and (d) (coloured by local 12-fold orientation) visually confirm these trends. At $\gamma_{\max}-\gamma_{\rm yield}\simeq 0.005\simeq 0$, the global orientation is well-defined, whereas at $\gamma_{\max}-\gamma_{\rm yield}=0.135$ the system fragments into large domains of differing orientation. For large $\gamma_{\max}$, strong deformation and frequent avalanches during each cycle locally disrupt orientational order. As a result, while $\langle q_{12}\rangle$ remains high within domains, its average over the whole system decreases due to the defects concentrated at domain boundaries.

The ‘‘best'' quasicrystals, as indicated by a high $\langle Q_{12}\rangle$,  occur near yielding but at the cost of a sharp increase in the \textit{timescales}: the number of cycles required for self-assembly grows sharply as $\gamma_{\max}\to\gamma_{\rm yield}$. This is shown in Fig.~\ref{fig:4}(e), where we plot the time it takes each system to reach the steady state. Nevertheless, for the sizes and density studied here, the assembly time stays modest (below $\sim 100$ cycles), and for $\gamma_{\max}-\gamma_{\rm yield}>0.1$ it is typically under five cycles. Thus, cyclic shear assembles high-quality quasicrystals far more rapidly (in terms of CPU time) than low-temperature molecular dynamics that rely on thermal fluctuations\textsuperscript{\dag}. Part of this acceleration arises from a weak orientational bias induced by the shear. As shown in the inset of Fig.~\ref{fig:4}(e), the orientation probability of the final quasicrystal is not uniformly distributed, but instead favours specific alignment angles relative to the shear direction. However, this effect is unlikely to be the only driver of the fast self-assembly of quasicrystals under shear, especially since this distribution changes little between $\gamma_{\max}-\gamma_{\rm yield}\simeq 0$ and $\gamma_{\max}-\gamma_{\rm yield}=0.1$. We believe that the more important mechanism is the system-spanning rearrangements triggered by plastic events, which enable large-scale exploration of configuration space, analogous to cluster moves in Monte Carlo simulations. This \textit{non-local} mechanism explores the energy landscape more effectively than the \textit{local} thermal motion of standard molecular dynamics simulations.

Another relevant timescale is the number of cycles required for a quasicrystal in the steady state to change its global orientation. This reorientation time is distinct from the assembly time from a random initial condition and, in equilibrium, would typically be very large because local dynamics cannot easily cross the free-energy barrier between differently oriented crystals in a box. Fig.~\ref{fig:4}(f) reports the typical timescale, in number of cycles $n_c$, extracted from the 12-fold orientation autocorrelation in the steady state,
\begin{equation}
    \begin{split}
    C_{12}^{\text{angle}}(n_c)&=\dfrac{\left\langle\tilde Q_{12}^{~}(n_c'+n_c)\tilde Q_{12}^*(n_c')\right\rangle}{\left\langle\left|\tilde Q_{12}(n_c')\right|^2\right\rangle},
    \end{split}
    \label{eq:C(t)}
\end{equation}
(shown in inset). The angular brackets denote a running average over multiple independent systems and over multiple $n_c'$ large enough to have reached a steady state. Although reorientation is slower than assembly, it remains finite, and no plateau appears in the correlation function. Quasicrystal reorientations are observed to proceed via nucleation of a domain with a different orientation, at least for small $\gamma_{\max}$.

\subsection{System size scaling}

\begin{figure}
\centering
  \includegraphics[width=0.485\textwidth]{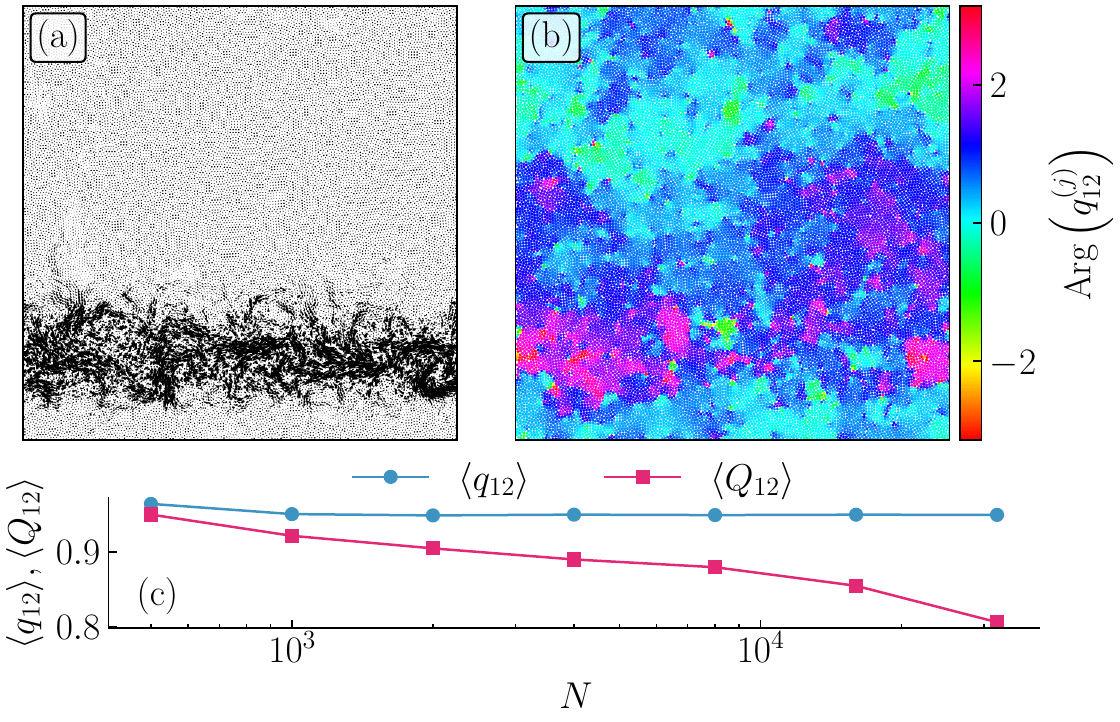}
  \caption{Quasicrystal quality dependence on the system size. (a) and (b) are snapshots of a large system ($N = 25000$) at $\rho\sigma^2 = 0.966$ and $\gamma_{\max}=0.07$. (a) displacement field (in arbitrary units of length) long after the band formation, between cycles 850 and 1000. (b) snapshot with particles coloured according to their 12-fold local orientation. (c) Evolution of the local and global BOOPs (Eq.~\eqref{eq:boop}) as $N$ is increased for $\gamma_{\max}=0.085$. Averages are performed over 3 realisations for 25 uncorrelated cycles after 1000 cycles for each system size.}
  \label{fig:5}
\end{figure}

Fig.~\ref{fig:2} showed that strong shear localisation (e.g., shear bands) can form in our system. For the system sizes used in cyclic shear up to now, this was not an issue: shear bands were highly mobile, appeared only at large $\gamma_{\max}$, and had a negligible impact on the dynamics. This is not true for larger systems. As shown in Fig.~\ref{fig:5}(a), which shows the displacement field between configurations 150 cycles apart for $N=25000$, a shear band forms clearly and persists over many cycles. In particular, for small $\gamma_{\max}$, the band forms around the 100$^{\rm th}$ cycle and remains localised thereafter. The transient state preceding band formation allows the system to relax into multiple quasicrystalline domains (Fig.~\ref{fig:5}(b)). However, once the shear band forms, plastic activity becomes confined to this region, preventing further equilibration of the domains outside the band. The tiling within the band itself typically develops a distinct orientation, different from its surroundings, further fragmenting long-range orientational coherence.

Increasing the system size, therefore, poses two key obstacles to achieving global quasicrystalline order: (i) in large systems, the shear band is intrinsically disordered, and (ii) because strain localises within the band, regions outside it cannot equilibrate globally on accessible timescales. This is quantified in Fig.~\ref{fig:5}(c). While the local order $\langle q_{12}\rangle$ is largely insensitive to $N$, since tiles form rapidly, the global order $\langle Q_{12}\rangle$, which captures coherent orientation across the system, decreases with increasing $N$. In principle, the shear band could eventually migrate and, in this way, contribute to the reorganisation of the entire system; such behaviour is not observed within feasible simulation times for $N > 15000$. Consequently, areas surrounding the band remain largely unaffected after the initial transient, further reducing $\langle Q_{12}\rangle$ in large systems.

Since strain localisation reduces the ability to develop global quasicrystalline order, as evidenced by the decrease of $\langle Q_{12}\rangle$ with system size, it is tempting to explore conditions under which the appearance of shear bands is strongly suppressed. This is precisely what we examine in the following section, where we show that such conditions provide only marginal improvement.

\subsection{Finite rate cyclic shear}

As discussed in the previous section, under the AQS protocol, pronounced strain localisation limits the formation of high-quality quasicrystalline order. We therefore turn to driving at a finite shear rate, which is known to promote more homogeneous deformation via many thin, mobile shear bands distributed throughout the sample~\cite{singh2020brittle}, and may thus circumvent the limitations of AQS.

To study finite-rate shearing, we use a co-shearing simulation cell in which particle coordinates are affinely remapped with the deforming box and evolve under overdamped dynamics:
\begin{equation} 
    \dfrac{d\gamma}{dt}=\dot\gamma(t), \qquad \Gamma \dfrac{d\mathbf r_i}{dt}=-\dfrac{\partial U}{\partial\mathbf r_i},
\end{equation}
where the imposed strain rate is constant $\dot\gamma(t)\equiv \dot\gamma$ except for sign reversal at $|\gamma|=\gamma_{\max}$ and $\Gamma$ is a friction coefficient. The box deformation therefore evolves as $\gamma(t)=\dot\gamma t$ between reversals, and the particle position equations are integrated simultaneously with the box deformation. In the limit $\Gamma\sigma^2\dot\gamma/\varepsilon\to 0$, the dynamics approach the AQS regime, since the system fully relaxes between successive plastic events. At larger $\Gamma\sigma^2\dot\gamma/\varepsilon$, the system may still be dissipating energy from one plastic event when the imposed shear triggers the next, so the time-scale separation that underlies AQS no longer holds. As before, we analyse the dynamics stroboscopically and, to enable direct comparison with the AQS results, we minimise the energy to obtain the inherent structure before measuring structural observables.

\begin{figure*}[!ht]
    \includegraphics[width=0.99\textwidth]{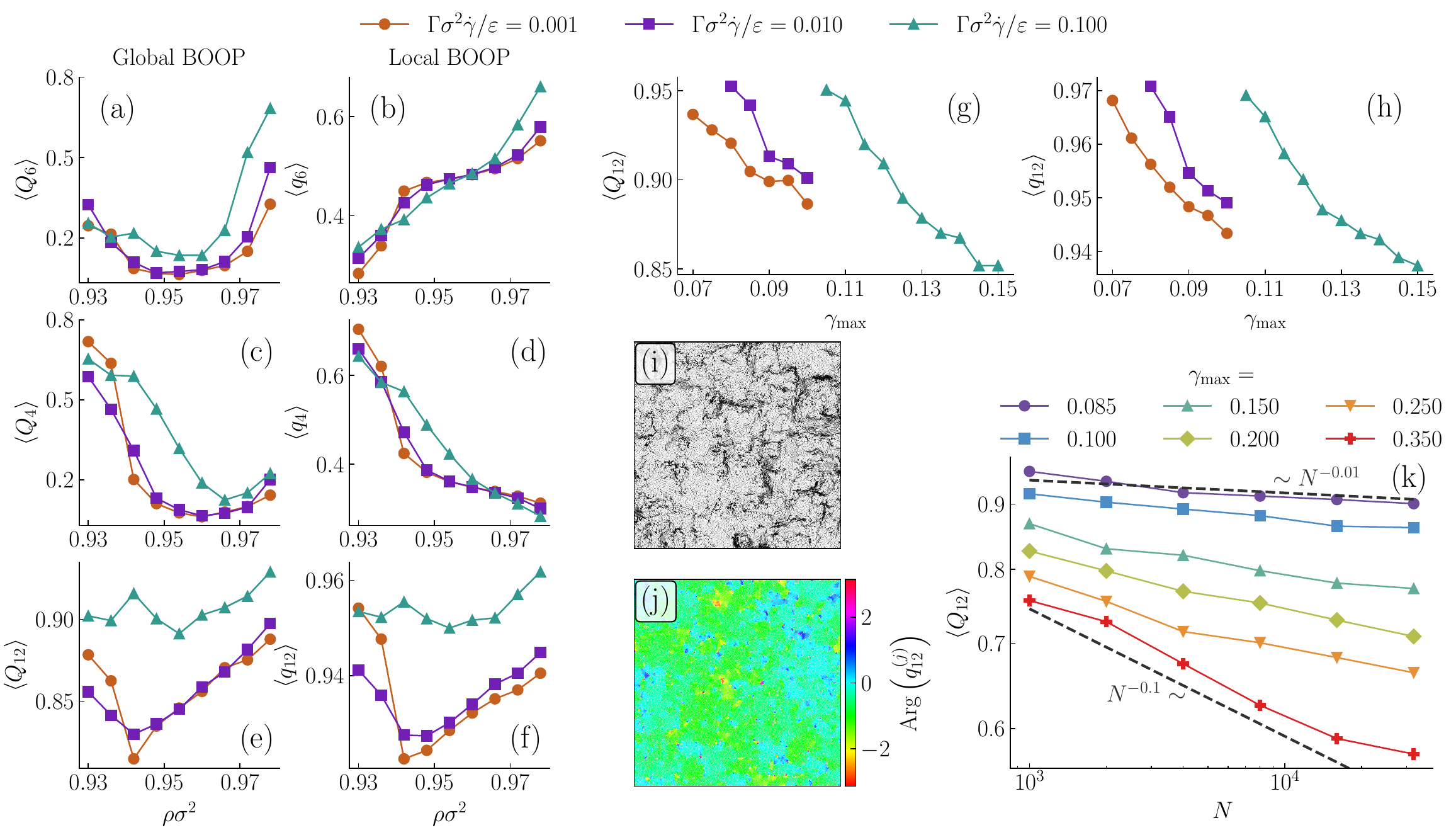}
  \caption{Behaviour of the system under cyclic finite shear rate. (a-f) Evolution of the global (left) and local (right) BOOPs as the density is varied. $\gamma_{\max}=0.12$ and $N=12000$. (g-h) Evolution of $\langle Q_{12}\rangle$ and $\langle q_{12}\rangle$ as $\gamma_{\max}$ is varied for three different $\dot \gamma$. For each curve, the minimum $\gamma_{\max}$ corresponds to the smallest value simulated before the system falls below yielding and undergoes the absorbing-state transition. $\rho\sigma^2=0.967$ and $N=2000$. (i-j) Displacement and orientation field (between 100 cycles) at $\rho\sigma^2=0.967$, $\Gamma\sigma^2\dot\gamma/\varepsilon=0.01$ and $N=32000$. (k) System size scaling of the global orientation order $\langle Q_{12}\rangle$ for various $\gamma_{\max}$. Each data point is an average over 3 different initial configurations and 10 to 40 uncorrelated snapshots.}
  \label{fig:6}
\end{figure*}

The simulation results are shown in Fig.~\ref{fig:6}. Panels (a–f) reproduce the AQS results for the density dependence of the local and global BOOPs at $\gamma_{\max}=0.12$ for three strain rates. In all cases, we observe the same trend: large $\langle q_{4}\rangle$ at low density, large $\langle q_{6}\rangle$ at high density, and a quasicrystalline phase in between. Fig.~\ref{fig:6}(g–h) focus on the single density $\rho\sigma^2 = 0.967$, where a quasicrystal is expected. For each $\dot\gamma$, the leftmost point corresponds to the smallest $\gamma_{\max}$ exceeding the observed $\gamma_{\rm yield}$, which itself depends on $\dot\gamma$. We observe again that the highest-quality quasicrystals appear around $\gamma_{\max}=\gamma_{\rm yield}(\dot\gamma)$. Although these observables mirror the AQS case, Fig.~\ref{fig:6}(i) shows that for $\Gamma\sigma^2\dot\gamma/\varepsilon=0.01$ and $N=32000$ the displacement map between multiple cycles is relatively homogeneous, with no sign of strain localisation for such $\dot\gamma$, contrary to AQS for a similar system size. The orientation map in Fig.~\ref{fig:6}(j) is likewise fairly homogeneous. We expect that a well-localised shear band should reappear as $\Gamma\sigma^2\dot\gamma/\varepsilon\to0$.

Finally, we examine the system-size dependence of $\langle Q_{12}\rangle$ in Fig.~\ref{fig:6}(k). Surprisingly, even under homogeneous driving, $\langle Q_{12}\rangle$ decreases with $N$. The decay is slow---making it difficult to discriminate logarithmic from power-law behaviour---but it suggests that orientational order is not truly long-ranged and is instead most likely quasi-long-range\textsuperscript{\dag}, with $\langle Q_{12}\rangle \overset{N\to\infty}{\longrightarrow} 0$. We cannot, however, exclude a finite plateau as in equilibrium, nor the possibility that true long-range order emerges precisely at $\gamma_{\max}=\gamma_{\rm yield}$.  This phenomenology is reminiscent of equilibrium hexatic or tetratic phases, yet at low temperature a 12-fold quasicrystal is expected to exhibit long-range order at equilibrium~\cite{dotera2014mosaic}. The observed ``dodecatic'' state therefore appears to be a genuinely non-equilibrium signature, indicating that the driving induces fluctuations large enough to disrupt order even for weak driving. By analogy with Kosterlitz--Thouless--Halperin--Nelson--Young melting, where the solid--hexatic transition is driven by dislocation unbinding (suppressed at low $T$ because dislocations are costly), cyclic driving may instead generate unbound dislocations in the quasicrystal and thereby destroy long-range orientational order. Consistent with this picture, the snapshot in Fig.~\ref{fig:6}(j) closely resembles configurations in hexatic phases: it shows an apparent global orientation (which would not persist in the thermodynamic limit) together with several misoriented domains~\cite{bernard2011two}. Interestingly, other non-equilibrium systems have been shown to exhibit enhanced fluctuations that destroy crystalline long-range order~\cite{PhysRevE.104.064605, PhysRevLett.131.108301}, whereas shear on scalar fields or cyclic driving is typically thought to suppress fluctuations and promote long-range order~\cite{giomi2022long, nakano2021long,  minami2021rainbow, ikeda2024dynamical, ikeda2024continuous, ikeda2024harmonic, kuroda2024longrange}.

\section{Conclusion}

We investigated the self-assembly of crystals and quasicrystals under cyclic shear and found that all phases that form at low temperature in equilibrium also emerge under athermal quasi-static cyclic driving. The non-local and orientation-biased nature of the driving enables remarkably fast self-assembly of the quasicrystal, whose quality peaks near the yielding amplitude, precisely where the dynamics slow down the most. To suppress strain localisation and the associated inhomogeneous driving, we turned to finite-rate shear. This approach preserves all the qualitative features observed under quasi-static driving while ensuring a more homogeneous deformation. We further showed that the resulting quasicrystalline order is not long-ranged, but instead quasi–long-ranged.

Several avenues remain open. A key question is the robustness of this self-assembly across interaction potentials, including for crystalline phases, whose behaviour under simple AQS protocols has also not been systematically explored. Toward this direction, obtaining a fully resolved $\rho$--$\gamma_{\max}$ phase diagram in the large system limit and comparing it with its equilibrium $\rho$--$T$ analogue would help clarify whether the same qualitative phase boundaries systematically emerge and whether the lever rule is generically violated.

It is also important to determine how quasicrystals formed under cyclic shear differ structurally from their equilibrium counterparts at a more microscopic level, for instance, in terms of defect topology, phason strain, or stress-structure correlations, which would require dedicated analyses beyond the scope of the present study.
The presence of residual stress indicates that they occupy distinct regions of the configurational landscape, even though they assemble at comparable densities.

The non-equilibrium driving allows for fast relaxation and even reorientation, suggesting that global rearrangements enable an effective exploration of the structural landscape. However, the mechanisms underlying the exploration of this landscape remain to be studied, such as the nucleation-like event observed before a global reorientation. Importantly, this dynamics offers a practical route to generating coherently oriented quasicrystalline domains, a task that is notoriously difficult to achieve in equilibrium and typically requires careful thermal annealing, which is often less accessible experimentally than controlling $\gamma_{\max}$.

Since the highest-quality quasicrystals arise near yielding, where dynamics are slowest, it would be interesting to explore strategies to accelerate their formation, such as linear annealing of $\gamma_{\max}$ or optimal-control protocols. At the same time, this correlation between slow dynamics and high structural quality raises a more fundamental question: what is the microscopic pathway by which cyclic shear drives quasicrystal formation? The fact that the best quasicrystalline order is obtained close to yielding, where the relevant time scales are also longest, suggests that slow relaxation may play a central role in selecting or stabilizing the ordered state. A natural next step would be to analyse the particle trajectories during assembly to identify how order develops---whether through progressive local defect removal, domain nucleation and coarsening, or genuinely system-spanning collective events. Such an analysis would address the origin of the fast assembly under shear and clarify why the vicinity of yielding appears to optimise quasicrystal quality.

Since for the AQS protocol, shear bands inevitably form for sufficiently large system sizes, we also explored quasicrystal formation at finite shear rates, where shear band formation is expected to be suppressed. We observed that our quasicrystals formed under finite-rate shear exhibit quasi–long-range orientational order. At equilibrium, such behaviour is associated with defect unbinding, and probing defect statistics and orientational correlations in these driven systems, particularly in the large-system limit, may reveal whether an analogous mechanism is at play, and whether the typical scenario observed in equilibrium 2D melting can be of any use to understand the observed quasi-long-range order.

Finally, it is natural to ask how these observations would extend to three dimensions, where the Kosterlitz–Thouless–Halperin–Nelson–Young scenario is absent and true long-range orientational order may be stabilised. It is therefore unclear whether cyclic shear can promote genuinely long-range quasicrystalline order, or whether it instead selects distinct non-equilibrium steady states.

\section*{Conflicts of interest}
There are no conflicts to declare.

\section*{Data availability}

The simulation scripts are publicly available at \href{https://doi.org/10.5281/zenodo.18254714}{https://doi.org/10.5281/zenodo.18254714}.

\balance

\bibliography{rsc}
\bibliographystyle{rsc}

\clearpage
\appendix

\renewcommand{\thefigure}{S\arabic{figure}}
\setcounter{figure}{0}

\renewcommand{\thetable}{S\arabic{table}}
\setcounter{table}{0}

\renewcommand{\theequation}{S\arabic{equation}}
\setcounter{equation}{0}

\onecolumn
\setcounter{page}{1}

\pagestyle{SIstyle}
\thispagestyle{SIstyle} 

\begin{center}
    \LARGE{\textbf{SUPPLEMENTARY INFORMATION}}
    
    \hspace{10pt}

    \large{Rapha\"el Maire, Andrea Plati, Frank Smallenburg and Giuseppe Foffi}

    \hspace{10pt}
\end{center}

\section{Details on numerical simulations}
\subsection{Forces, Potential and Box}
\begin{figure}[!h]
    \centering
    \includegraphics[width=0.8\linewidth]{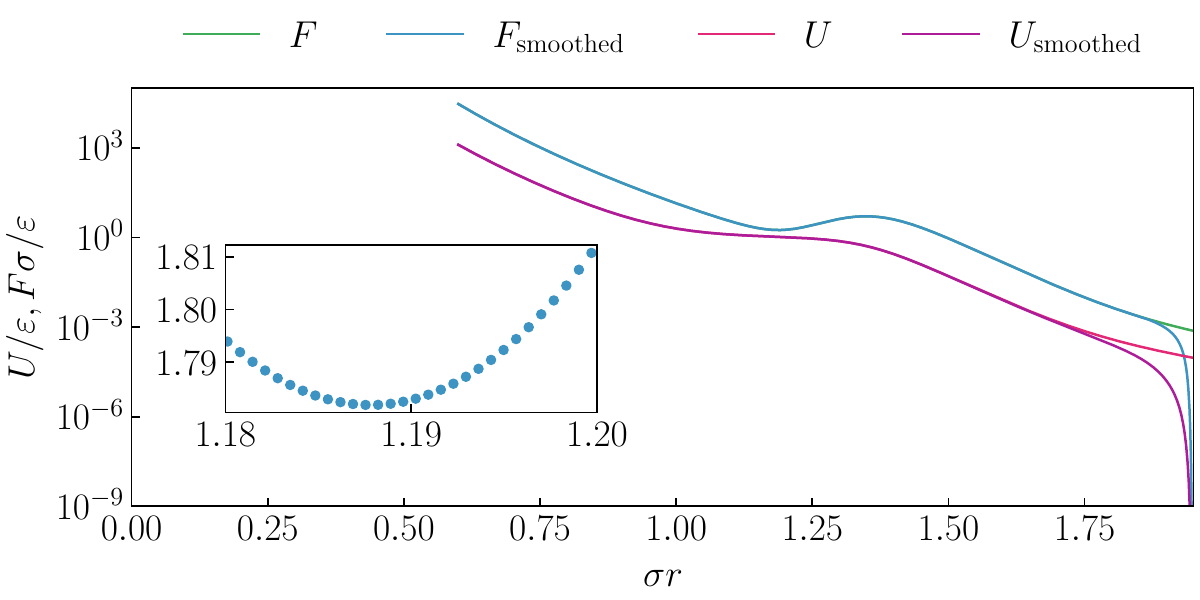}
    \caption{Smoothed forces and potential. In the inset, we show the typical sampling of points used for the tabulated force.}
    \label{fig:forcesmooth}
\end{figure}

Using \texttt{LAMMPS}, we simulate a deformable, periodic, two-dimensional box of constant height and width $L$. The box undergoes simple shear characterised by a shear strain $\gamma$ along the $x$ direction, resulting in a tilted simulation cell with lattice vectors:
\begin{equation}
    \bm{a}_x = (L, 0), \quad \bm{a}_y = (\gamma L, L) .
\end{equation}
$N$ particles of mass $m$ are placed in the box, which defines the density: $\rho = N/L^2$. Particles interact through a smoothed force of the form:
\begin{equation}
    F_{\rm smoothed}(r) = \left(-\dfrac{\mathrm{d}U}{\mathrm{d}r}\right) S_{\rm mollifier}(r; r_{\rm thres}, r_\mathrm{cut}),\qquad \dfrac{U(r)}{\varepsilon}=\left[ \left( \dfrac{\sigma}{r} \right)^{14} + \dfrac{1}{2}\left(1 - \tanh \left(  \dfrac{r-\delta}{w}\right)\right) \right]
\end{equation}
where $S_{\rm mollifier}(r; r_{\rm thres}, r_\mathrm{cut})$ is a mollifier defined as:
\begin{equation}
    S(r; r_{\rm thres}, r_\mathrm{cut}) =
    \begin{cases} 
    1, & r \le r_{\rm thres},\\
    \displaystyle \exp\left[1-\frac{1}{1 - \left(\frac{r - r_{\rm thres}}{r_\mathrm{cut}-r_{\rm thres}}\right)^2}\right] , & r_{\rm thres} \leq r \leq r_\mathrm{cut},\\
    0, & r \ge r_\mathrm{cut}.
    \end{cases}
\end{equation}
This smoothing procedure ensures that both the force $F_{\rm smoothed}$ and the potential $U_{\rm smoothed}(r)\equiv\int_{r}^{r_{\rm cut}}F_{\rm smoothed}(r')\mathrm dr'$ vanish smoothly to 0 at the cut-off $r_{\rm cut}$, starting from $r_{\rm thres}$. Throughout our simulations, we chose $w=0.1\sigma$, $\delta = 1.35\sigma$, $r_{\rm thres}=1.85\sigma$ and $r_{\rm cut}=1.95\sigma$. The original and smoothed potentials and forces are shown in Fig.~\ref{fig:forcesmooth}. To facilitate efficient computation in \texttt{LAMMPS}, we tabulate the force and potential on 2000 evenly spaced points between $r=0.6\sigma$ to $r=r_{\rm cut}=1.95\sigma$. During the simulation, the values are obtained via spline interpolation using the \texttt{pair\textunderscore style table spline 2000} command. The sampling is illustrated in the inset of Fig.~\ref{fig:forcesmooth}.

The stresses are measured in \texttt{LAMMPS} via the virial stress.

\subsection{Equilibrium thermalisation}

For simulations requiring an initial thermalised configuration, we equilibrate the system at a target temperature $T$ using a Langevin thermostat with \texttt{LAMMPS} damping coefficient $0.1\sqrt{m\sigma^2/\varepsilon}$ and timestep $0.001\sqrt{m\sigma^2/\varepsilon}$. The system is annealed from an initial temperature $k_BT/\varepsilon=1$, which is linearly decreased to the desired value $T$. The annealing is performed over $10^7$ timesteps for systems with $N\leq 2\times10^4$, and $10^8$ timesteps for larger systems. The system is then held at the target temperature for an additional $10^7$ timesteps. We verified that both the potential energy and the pressure reach statistically stationary values following this procedure.

\subsection{Athermal quasi-static shear (AQS)}

Starting from a given configuration in an undeformed square box, we perform athermal quasi-static shear (AQS). Each AQS step consists of an affine shear increment $\gamma\to\gamma+\Delta\gamma$ with $\Delta\gamma = 10^{-4}$ followed by energy minimisation via Conjugate Gradient descent with a force tolerance of $10^{-10}\varepsilon/\sigma$. This two-step procedure is repeated until the box strain reaches $\gamma_{\max}$. For cyclic shear, once $\gamma_{\max}$ is reached, we reverse the sign of $\Delta \gamma$ and continue the simulation.

If $\gamma_{\max}/\Delta\gamma$ is not an integer, the deformation may slightly exceed $\gamma_{\max}$; in such cases, we reset the strain to exactly $\gamma\to\gamma_{\rm max}$ and re-minimise the energy. Similarly, when the strain changes from negative to positive values, we reset the strain to $\gamma\to 0$ to ensure that the system is analysed in an exactly undeformed box.

\subsection{Finite shear rate}

For simulations involving a finite shear rate, we perform overdamped dynamics using the \texttt{Brownian} fix of \texttt{LAMMPS} at zero temperature:
\begin{equation}
    \dfrac{d\gamma}{dt}=\dot\gamma \qquad \Gamma \dfrac{d\mathbf r_i}{dt}=-\dfrac{\partial U}{\partial \mathbf r_i},
\end{equation}
where $\dot\gamma$ is the imposed strain rate. The box deformation thus evolves according to $\gamma(t)=\dot\gamma t$ (in between strain-rate reversal events), and the particle equations of motion are integrated simultaneously with the box shear. The timestep is set to $0.001\Gamma \sigma^{2}/\varepsilon$.

\section{Ground state phase diagram}

\subsection{Energy of each phase}

\begin{figure}[!h]
    \centering
    \includegraphics[width=0.8\linewidth]{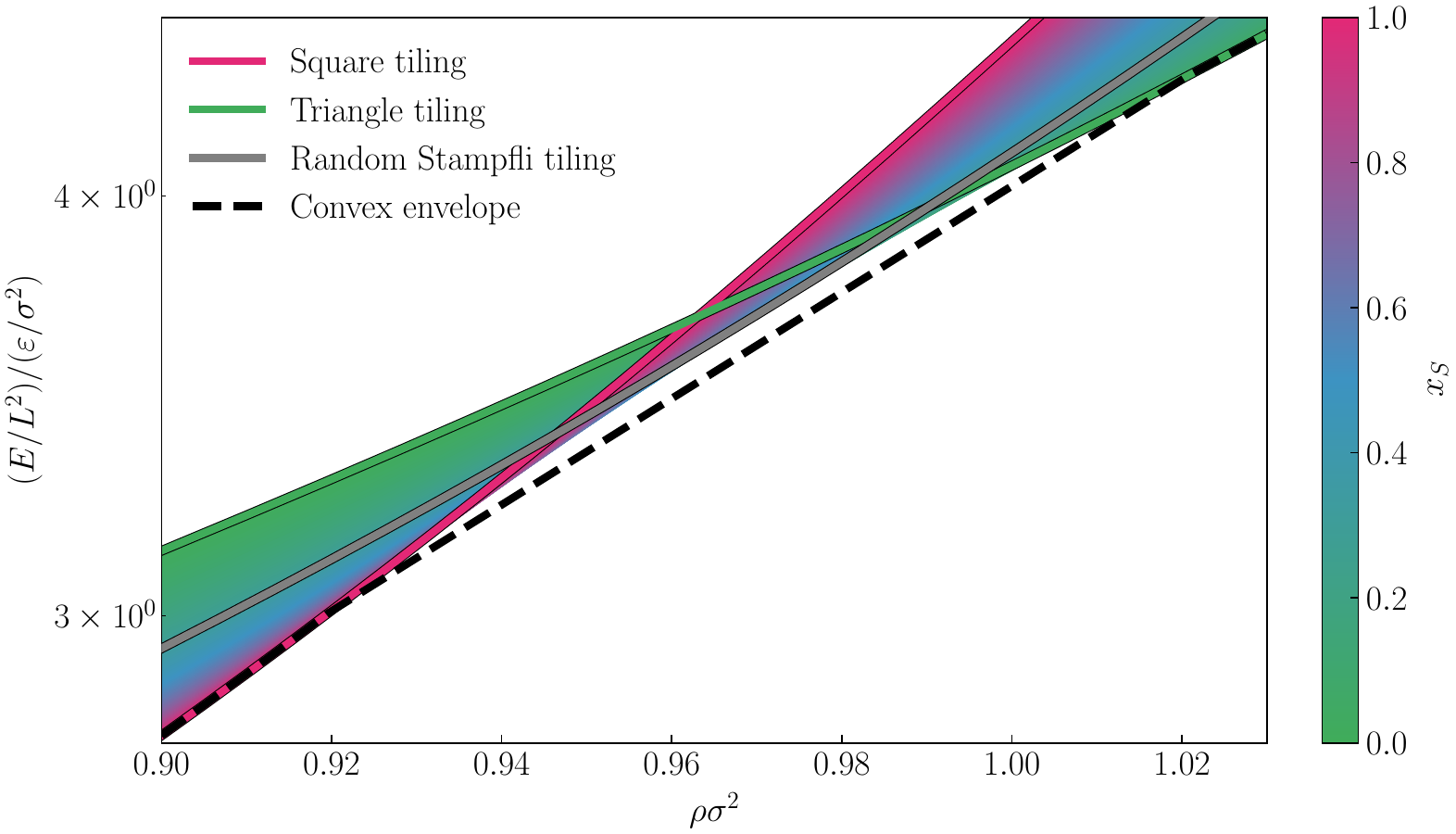}
    \caption{Energy of the square, triangle, Stampfli, and random  ‘‘mean-field'' (in shade of colour) tiling as a function of the density. The lowest energy state is never a random tiling. The y-axis is logarithmically scaled for better visualisation.}
    \label{fig:stability}
\end{figure}

The $T=0$ phase diagram can be obtained straightforwardly since the free energy density $f=(E-TS)/L^2$ reduces simply to the ground-state energy density. We compare the energies of the square and hexagonal lattices in Fig.~\ref{fig:stability} and determine the coexistence densities using a Maxwell construction. In the same figure, we also verify that a random Stampfli tiling~\cite{zeng2006inflation}, or a random “mean-field” tiling (as described below) with a fraction $x_S$ of squares and $1-x_S$ of triangles, always has a higher energy than the square-hexagonal coexistence. Therefore, such tilings do not represent the stable phase, contrary to the claim made in Ref.~\onlinecite{kryuchkov2018complex}. While this does not exclude the possibility that a more complex quasicrystalline structure could be stable at $T=0$, it provides strong evidence that, at equilibrium, the quasicrystal is instead stabilised by entropy.

We now clarify what we mean by a “random mean-field” tiling.

\subsection{Random ‘‘mean-field'' tiling}

We compute the typical energy of a random tiling in the mean-field limit. By mean-field, we mean that each particle is assumed to experience an average (possibly non-integer) number of neighbours, that the positions of triangles and squares are uncorrelated, and that each nearest neighbour bond is the same length $s$. This excludes coexistence, where triangle and square regions would have different bond lengths.

Let $f_T$, $f_S$, and $f=f_S + f_T$ denote the number of triangles, squares, and total tiles per unit of area, respectively. The fraction of square and triangle tiles is given by:
\begin{equation}
    x_S = \dfrac{f_S}{f_T + f_S}\quad\text{and}\quad x_T = 1-x_S.
\end{equation}
The areas of square and triangle tiles are obtained from their edge length $s$:
\begin{equation}
    A_S = s^2\quad\text{and}\quad A_T = \dfrac{\sqrt 3}{4}s^2.
\end{equation}
Since we perfectly tile the plane, we must have:
\begin{equation}
    f_TA_T+f_SA_S = 1\Rightarrow(1 - x_S)f\dfrac{\sqrt{3}}{4}s^2+x_Sfs^2=1,
    \label{eq:unknownfs}
\end{equation}
where $f$ and $s$ are the unknown.
\paragraph*{Determination of $f$:} The number of edges per unit area $e$ is given by:
\begin{equation}
    e = \dfrac{3f_T+4f_S}{2}=\dfrac{3 + x_S}{2}f
\end{equation}
where the factor $2$ comes from the fact that each edge is shared by two tiles. We now use Euler's identity, using the fact that the number of vertices is equal to the particle density:
\begin{equation}
    \rho - e + f = 0\Rightarrow f = \dfrac{2\rho}{1 + x_S}
    \label{eq:euler}
\end{equation}
\paragraph*{Determination of $s$:} Replacing Eq.~\eqref{eq:euler} into Eq.~\eqref{eq:unknownfs} leads to an expression for the edge length:
\begin{equation}
    s = \sqrt{\dfrac{1+x_S}{2\rho\left(x_S + (1 - x_S)\frac{\sqrt 3}{4}\right)}},
\end{equation}
which interpolates smoothly between the known result $\rho s^2=\dfrac{2}{\sqrt 3}$ for the hexagonal lattice ($x_S=0$) and $\rho s^2=1$ for the square lattice ($x_S=1$). 

\paragraph*{Determination of the number of closest neighbours:} To obtain the energy of the system at the \textit{mean-field level}, we must find the \textit{average} number of closest neighbours $z$ per vertex, which are at distance $s$. Since each edge connects two vertices, we find:
\begin{equation}
    z = \dfrac{2e}{\rho}=\dfrac{2(3+x_S)}{1 + x_S},
\end{equation}
which interpolates smoothly between $z=4$ for a square lattice and $z= 6$ for a hexagonal lattice. 
\paragraph*{Determination of the number of second-closest neighbours:} Since the energy cut-off is at $1.95\sigma$, we need to consider second-nearest neighbours. We will neglect the one arising from the triangles that lie at a distance $\sqrt 3s$ (which should be negligible for densities below $\rho\sigma^2\simeq 1$ since $U(1.75\sigma)\simeq10^{-3}\varepsilon$) and only consider the ones arising from squares, lying at $\sqrt{2}s$. Since each square has two diagonals with two particles each, the number of second-closest neighbours $z_2$ per vertex is:
\begin{equation}
    z_2=4\dfrac{f_S}{\rho}=\dfrac{8x_S}{1+x_S},
\end{equation}
which vanishes in the absence of squares and reaches $4$ for a perfect square lattice.

\paragraph*{Determination of the potential energy:} The potential energy is:
\begin{equation}
    E = \dfrac{1}{2}zU(s) + \dfrac{1}{2}z_2U(\sqrt{2}s).
\end{equation}
We neglected contributions beyond second-nearest neighbours and assumed that each particle experiences, on average, $z$ neighbours and $z_2$ second-closest neighbours (note that $z$ and $z_2$ do not have to be integers).

\section{Self-assembly speed}

\begin{figure}[!h]
    \centering
    \includegraphics[width=0.9\linewidth]{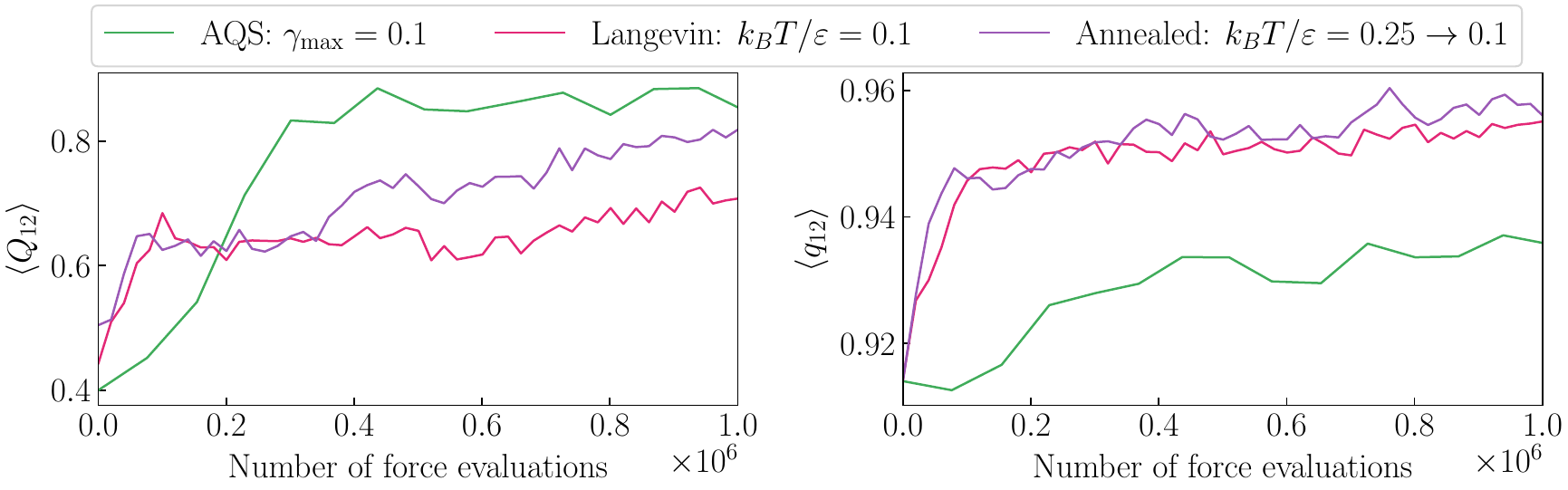}
    \caption{Self-assembly speed, measured via the local and global 12-fold bond-orientational parameters, in terms of the number of force evaluations. All systems are taken at $\rho\sigma^2=0.95$ and $N=3000$. ‘‘AQS'' is a system simulated via the quasistatic sheared protocol described in the main text at $\gamma_{\rm max}=0.1$,  ''Langevin'' is a system self-assembled via a Langevin bath at equilibrium and temperature $k_BT/\varepsilon=0.1$ and ‘‘Annealed'' is a system self-assembled by decreasing linearly the temperature with time, from $k_BT/\varepsilon=0.25$ to $0.1$.}
    \label{fig:q_12speed}
\end{figure}

In Fig.~\ref{fig:q_12speed}, we report the typical self-assembly time, measured in terms of the number of force evaluations, for three cases: an AQS protocol, a system thermalized with a Langevin bath, and a thermally annealed system. We find that the AQS protocol typically leads to faster quasicrystal self-assembly than the thermal protocol, for the chosen target temperature.

\section{System size scaling of \texorpdfstring{$\langle Q_{12}\rangle$}{Q12}}

To obtain the scaling of $\langle Q_{12}\rangle$ with system size, it is easier to deal with:
\begin{equation}
    \langle Q_{12}^2\rangle\equiv \left\langle \dfrac{1}{N^2}\sum_{j\neq k}q_{12}^{(j)}q_{12}^{*(k)}\right\rangle.
\end{equation}
$\langle Q_{12}\rangle$ should behave roughly as $\sqrt{\langle Q_{12}^2\rangle}$. We define the correlation function:
\begin{equation}
    C_{12}(r) = \frac{\displaystyle \left\langle \sum_{j\neq k} q_{12}^{(j)} q_{12}^{*(k)}\delta(r - |\mathbf r_j - \mathbf r_k|)\right\rangle}{\displaystyle \left\langle \sum_{j\neq k} \delta(r - |\mathbf r_j - \mathbf r_k|)\right\rangle}.
\end{equation}
To obtain the continuum form, we perform some manipulation on the following:
\begin{equation}
\begin{aligned}
 \langle Q_{12}^2\rangle=\dfrac{1}{N^2}\left\langle \sum_{j\neq k} q_{12}^{(j)} q_{12}^{*(k)} \right\rangle
&= \left\langle \sum_{j\neq k} \int_0^L  q_{12}^{(j)} q_{12}^{*(k)} \delta(r - r_{jk})dr \right\rangle \\
&= \int  \left\langle \sum_{j\neq k} q_{12}^{(j)} q_{12}^{*(k)} \delta(r - r_{jk}) \right\rangle dr \\
&= \int  C_{12}(r)
\left\langle \sum_{j\neq k} \delta(r - r_{jk}) \right\rangle dr\\
&=\dfrac{\rho}{N}\int_{[0, L_x]\times[0, L_y]}C_{12}(r)g(r)d^2\mathbf r.
\end{aligned}
\end{equation}
Where we recognised the pair-correlation function $g(r)$. At large distances, $g(r)\simeq 1$, therefore, we obtain, for large $L$, the scaling of $\langle Q_{12}\rangle$:
\begin{equation}
    \langle Q_{12}\rangle\sim\sqrt{\dfrac{1}{N}\int_0^{L} rC_{12}(r)dr}\sim\left\{
    \begin{array}{ll}
        N^0 & \mbox{if}\quad C_{12}(r\to\infty)=c_0 \\
        N^{-\eta/4}&\mbox{if}\quad C_{12}(r\to\infty)\sim r^{-\eta}\mbox{ with } 
        0<\eta<2 \\
        N^{-1/2} & \mbox{if}\quad C_{12}(r\to\infty)\sim e^{-r/\xi} \mbox{ or } C_{12}(r\to\infty)\sim r^{-\eta'} \mbox{ with } \eta'>2
    \end{array}
\right.
\end{equation}

The $N^0$ scaling corresponds to a system with a well-defined orientation in the infinite system size limit. In this case, it is orientationally long-range ordered. The $N^{-\eta/4}$ scaling describes a slow decrease of global orientation with system size. The orientational order is only quasi-long ranged. In the last case, the orientational order is lost as $N^{-1/2}$, which is the central limit theorem; we say that the orientational order is short-ranged (note that this scaling can also be obtained with a strong power-law decay of the correlation function).

\end{document}